\documentstyle[12pt,epsf]{article}
\begin{document}
\begin{titlepage}
\pagestyle{empty}
\baselineskip=21pt
\rightline{Alberta Thy-9-93}
\rightline{February 1993}
\vskip .2in
\begin{center}
{\large{\bf Toponium Tests Of Top-Quark Higgs Bags}}
\end{center}
\vskip .1in
\begin{center}
Alick L. Macpherson and Bruce A. Campbell

{\it Department of Physics, University of Alberta}

{\it  Edmonton, Alberta, Canada T6G 2J1}

\vskip .2in

\end{center}
\centerline{ {\bf Abstract} }
\baselineskip=18pt

\noindent
Recently it has been suggested that top quarks, or very massive fourth
generation quarks, might surround themselves with a Higgs ``bag" of deformation
of the Higgs expectation value from its vacuum magnitude. In this paper we
address the question of whether such nonlinear Higgs-top interaction effects
are subject to experimental test. We first note that if top quarks were
necessarily accompanied by Higgs ``bags", then top quark weak decay would
involve the sudden disruption of the Higgs ``bag", with copious production of
physical Higgs particles accompanying the decay. We then examine the effects
that such Higgs ``bags" would produce on the spectrum of toponium, where the
two bound top (anti)quarks, and their ``bags", overlap. We numerically evaluate
the effects that the nonlinear feedback in the Higgs-toponium system would have
on the energy level splittings of the toponium bound states, and find that for
allowed values of the top and Higgs mass the effect is negligible, thus
indicating that even in this favourable circumstance Higgs ``bag" formation
around top quarks does not observably occur. Finally, we consider the case of a
second Higgs doublet, allowing the possibility of enhanced couplings for one of
the physical Higgs to top. Even in this nonstandard scenario the effects are
minimal, and we infer the general absence of observable effects at any level
that might suggest the utility of considering top quarks to be accompanied by
Higgs ``bags".
\end{titlepage}
\baselineskip=18pt

In the standard model the Higgs field acts, through its vev, as the source of
mass for all particles, with the mass obtained depending on the strength of the
particle's coupling to the Higgs. Of the particles in the standard model, the
only one with potentially very large mass, and hence large coupling to the
Higgs, is the top quark. This opens the possibility that there are
nonperturbative, strong-coupling effects, with Higgs particles, that will occur
uniquely in interaction with the top quark. The idea that a fermion which is
strongly coupled to an order parameter may locally deform that order parameter,
and surround itself with a ``bag" of field deformation, dates at least as far
back as Feynman's treatment of polarons \cite{fey}, and more recently has been
generally explored in relativistic field theories of scalars and spinors
\cite{vin,lw,bdwy,fl,gw,mwz,mac}. Recently it has been suggested \cite{wil}
that for large values of the top quark mass just such nonlinear effects occur,
with the top quark digging a hole in the Higgs vev, and surrounding itself with
a ``bag" or ``dimple" of deformation of the Higgs field (a posteriori such a
possibility would also appear for very massive quarks, or leptons, of a
hypothetical fourth generation). More detailed quantitative examinations of
this proposal have come to the conclusions that: semiclassical ``bag" formation
implies couplings sufficiently strong to jeapordize vacuum stability, or imply
a breakdown of perturbation theory at energies not too far above the top quark
mass range \cite{dim}; perturbative couplings result in ``dimples" that as
quantum  superpositions involve on average a fraction of a quantum \cite{and};
strong non-perturbative couplings result in quantum fluctuations that tend, at
least in a large N expansion, to ``deflate" the ``bag" \cite{bag}. In this
paper we adopt a slightly different approach to the problem; we ask what would
be the observable signatures of formation of Higgs ``bags", both for individual
top quarks, and also for toponium bound states. We then evaluate the magnitude
of these effects for top quarks of moderate mass, where we may treat the
Higgs-top coupling in perturbation theory, and examine where the nonlinear
higher-order effects should begin to dominate, giving observable signatures of
``bag" formation. In agreement with the previous analyses \cite{dim,and,bag} we
find for standard model Higgs masses in the range allowed by vacuum stability,
and perturbative non-triviality, that the effects of Higgs ``bag" formation are
not strong enough to be significant. We then extend our analysis to the case of
two Higgs doublets, where one of the Higgs may have enhanced coupling to the
top, to examine whether in this case observable effects of Higgs ``bag"
formation may occur.

The possibility of the formation of Higgs ``bags" around heavy quarks is
suggested by simple energetic considerations. A heavy quark obtains its large
mass by virtue of a large Yukawa coupling to the Higgs field vev. If the value
of that vev could be locally diminished in the vicinity of the top quark, then
the mass of the top quark could be lowered. Provided that the gain in energy
from decreasing the mass of the top quark can more than compensate for the
kinetic and potential energy invested in deforming the Higgs field around the
top quark, and the kinetic energy localizing the top quark, then the top quark
will dig a hole for itself in the Higgs vev, and inhabit the region of
diminished vev. For this to be energetically favourable, we need the energy
saved from lowering the quark mass to dominate, which means the possibility
depends on a large Yukawa coupling, and so it may occur only for (very) heavy
quarks. If this scenario is correct, then a heavy quark such as the top should
be thought of not as an isolated fermion, but rather as a structured object
consisting of a fermion surrounded by a coherent superposition of Higgs bosons
representing the deformation of the Higgs vev.

Since this coherent superposition of Higgs quanta is supported by the energy
saved in reducing the mass of the heavy quark source, the disappearance of that
quark would of necessity result in the dispersal of the Higgs quanta. In the
case of top quarks, this means that their normal charged current weak decay,
via $t \rightarrow bW^{+}$ would remove the source of the Higgs ``bag" (the $b$
being too weakly coupled to the Higgs), and hence lead to the sudden disruption
of the ``bag". This would in turn mean that the dominant decay modes of such a
top quark (with ``bag") would involve a copious shower of Higgs bosons from the
disruption of the ``bag", as well as the $b$ and $W$. Decay to the $bW$ mode
(without Higgs) would be strongly suppressed by the small wave function overlap
of the ``bag" state with the final state absence of Higgs. This means that the
observation of the standard decay mode of the top would provide prima facie
evidence against the formation of Higgs ``bags". Conversely, a fermion strongly
enough coupled to the Higgs field to engender ``bag" formation, may be expected
to have complex decay modes, that display the complexity of the coherent Higgs
superposition in which it reposes.

A second way that one might imagine obtaining experimental evidence concerning
the possibility of Higgs ``bag" formation, is by examining toponium bound
states. A priori, these seem like ideal systems to probe the possibility of
``bags": first they represent already localized top quark sources for the
Higgs; second the bound state spectrum provides a sensitive test of the
structure of the potential well in which the $\bar{t}t$ find themselves, and
should surely be sensitive to as qualitatively distinct a feature as Higgs
``bag" formation. To reduce the problem to its essential form, let us consider
a $\bar{t}t$ bound state, held together by the QCD potential, which for the
heavy toponium we may consider to be approximately Coulombic, and which
interacts with the Higgs field via a Lagrangian of the form: (we ignore
everything else in the standard model, as we expect it to be quantitatively
insignificant in our considerations)
\begin{equation}
{\cal L} =\frac{1}{2} \partial_{\mu}\phi \partial^{\mu}\phi -
\frac{m_{H}^{2}}{2} \phi^2 + \bar{\Psi}(i\not\!\partial - g \phi) \Psi -
m_{t}\bar{\Psi} \Psi - \frac{\lambda}{4!} \phi^{4}
\label{1}
\end{equation}
The interaction term ${\cal L} =  - g \bar{\Psi} \phi \Psi$ will cause a minor
deformation of the Higgs field in the presence of the top quarks. Moreover, due
to the assumed the heaviness of top, one can apply  non-relativistic
quasi-classical methods to bound state systems (toponium)composed of (anti)top
quarks and a Higgs field.  For quasi-static (anti)top quarks in a toponium
bound state, which act as a source of deformation of the Higgs field from its
vev, we have classically for the Higgs deformation:
\begin{equation}
(\nabla^{2} - m_{H}^{2})\phi = g  \psi^{\dagger}\psi
\label{2}
\end{equation}
where in the preceding equation (and hereafter) the $\psi$ represents the
``large" components of the non-relativistic spinor $\Psi$.
Here the time dependence has been disregarded as the lowest energy state of the
toponium is stationary. Further, the scalar coupling of the Higgs to the top
quark, and the non-relativistic treatment of the toponium, implies that the
spin degrees of freedom of the top can be neglected, and the ${\psi}^{\dagger}
\psi$ can be treated as a scalar source for the Higgs vev deformation.

We consider the S-wave fermion wave functions of our toponium bound states as
Higgs sources. In view of the spherical symmetry of the S-wave states, the
source term composed of the $\bar{t}t$ can be written in terms of the top wave
function, expressed in polar coordinates, centred on the toponium. Assuming
that the QCD binding potential is approximately Coulombic, then over the
distance scale probed by the toponium wave function, these wave functions are
exponential in nature; they act as an exponentially falling (radially) Higgs
source term; and the 1S and the 2S wave functions represent strong, localized
Higgs sources. For Coulombic toponium, the  Higgs field source terms are:
\begin{equation}
\begin{array}{lcrrcl}
1S:\hspace{2cm}   & &g {\psi}^{\dagger}\psi &=& \frac{g}{\pi a_{0}^{3}}
e^{-\frac{2r}{a_{0}}} \\
2S:\hspace{2cm}   &  &g {\psi}^{\dagger}\psi &=& \frac{g}{32 \pi a_{0}^{3}} ( 2
- \frac{r}{a_{0}}) e^{-\frac{r}{a_{0}}}
\end{array}
\label{4}
\end{equation}
with $a_{0} = \frac{1}{2 \pi \alpha_{s} m_{t}}$  as the Bohr radius of the
unperturbed toponium bound state. $\alpha_{s}$ is the effective strong coupling
constant on scales corresponding to the size of the toponium bound state,
which we take to have a value of $\alpha_{s}\simeq 0.32$, in approximate
agreement with the values used by Athanasiu et al. \cite{ath} in their study of
the $\bar{t}t$ system. The deformation of the Higgs field in the neighborhood
of the toponium source causes a decrease in the observed top (and hence
toponium) mass.

To solve for the deformation of the Higgs vev, one uses the three dimensional
Green's function associated with the equation of motion of the Higgs field.
\begin{equation}
G(\tilde{r}_{1},\tilde{r}_{2}) = \frac{-1}{4 \pi} \frac{e^{m_{H}|\tilde{r}_{1}
- \tilde{r}_{2}| }}{|\tilde{r}_{1} - \tilde{r}_{2}|}
\label{5}
\end{equation}
Utilising this, one then can then analytically obtain the first order position
dependent deviation  of the Higgs vev from its asymptotic value of $v=246$ GeV.
For the 1S and 2S toponium wave function sources the form of the Higgs vev
deviation is
\begin{equation}
\phi_{1S}^{1st}(r) ={{{\alpha_{s}^3}\,{m_{t}^3}\,\left(
2\,\alpha_{s}\,{e^{m_{H}\,r}}\,m_{t} -
       2\,\alpha_{s}\,{e^{\alpha_{s}\,m_{t}\,r}}\,m_{t} -
{e^{m_{H}\,r}}\,{m_{H}^2}\,r +
       {\alpha_{s}^2}\,{e^{m_{H}\,r}}\,{m_{t}^2}\,r \right) }\over
   {8\,{e^{\left( m_{H} + \alpha_{s}\,m_{t} \right) \,r}}\,
     {{\left( m_{H} - \alpha_{s}\,m_{t} \right) }^2}\,
     {{\left( m_{H} + \alpha_{s}\,m_{t} \right) }^2}\,\pi \,r}}
\label{7}
\end{equation}
\begin{eqnarray}
\phi_{2S}^{1st}(r)&=&{{\alpha_{s}\,{e^{-\left( m_{H}\,r \right)  -
{{\alpha_{s}\,m_{t}\,r}\over 2}}}\,
     m_{t}\,}\over
   {512\,{{\left( -2\,m_{H} + \alpha_{s}\,m_{t} \right) }^4}\,\pi \,r}}
\left( -128\,{e^{m_{H}\,r}}\,{m_{H}^3} +
       128\,{e^{{{\alpha_{s}\,m_{t}\,r}\over 2}}}\,{m_{H}^3} \right.  \nonumber
\\
       & & \left. +
       256\,\alpha_{s}\,{e^{m_{H}\,r}}\,{m_{H}^2}\,m_{t}-
       256\,\alpha_{s}\,{e^{{{\alpha_{s}\,m_{t}\,r}\over 2}}}\,{m_{H}^2}\,m_{t}
-
       128\,{\alpha_{s}^2}\,{e^{m_{H}\,r}}\,m_{H}\,{m_{t}^2}  \right.
\nonumber \\
       & & \left. +
       128\,{\alpha_{s}^2}\,{e^{{{\alpha_{s}\,m_{t}\,r}\over
2}}}\,m_{H}\,{m_{t}^2} +
       64\,{\alpha_{s}^3}\,{e^{m_{H}\,r}}\,{m_{t}^3} -
       64\,{\alpha_{s}^3}\,{e^{{{\alpha_{s}\,m_{t}\,r}\over 2}}}\,{m_{t}^3} -
       64\,\alpha_{s}\,{e^{m_{H}\,r}}\,{m_{H}^3}\,m_{t}\,r \right.   \\
       & & \left. +
       64\,{\alpha_{s}^2}\,{e^{m_{H}\,r}}\,{m_{H}^2}\,{m_{t}^2}\,r -
       64\,{\alpha_{s}^3}\,{e^{m_{H}\,r}}\,m_{H}\,{m_{t}^3}\,r  +
       24\,{\alpha_{s}^4}\,{e^{m_{H}\,r}}\,{m_{t}^4}\,r  \right.  \nonumber \\
       & & \left. +
       16\,{\alpha_{s}^2}\,{e^{m_{H}\,r}}\,{m_{H}^3}\,{m_{t}^2}\,{r^2}  -
       12\,{\alpha_{s}^4}\,{e^{m_{H}\,r}}\,m_{H}\,{m_{t}^4}\,{r^2} +
       4\,{\alpha_{s}^5}\,{e^{m_{H}\,r}}\,{m_{t}^5}\,{r^2}  -
       8\,{\alpha_{s}^3}\,{e^{m_{H}\,r}}\,{m_{H}^3}\,{m_{t}^3}\,{r^3} \right.
\nonumber \\
       & & \left. +
       12\,{\alpha_{s}^4}\,{e^{m_{H}\,r}}\,{m_{H}^2}\,{m_{t}^4}\,{r^3} -
       6\,{\alpha_{s}^5}\,{e^{m_{H}\,r}}\,m_{H}\,{m_{t}^5}\,{r^3} +
       {\alpha_{s}^6}\,{e^{m_{H}\,r}}\,{m_{t}^6}\,{r^3} \right)  \nonumber
\label{8}
\end{eqnarray}

To see the effect of the  coupling on the mass of the toponium bound state, we
examine the change in the splitting between the 2S and 1S energy levels; we
focus on the energy level splitting as it is a physical observable, and may
reasonably be expected to be sensitive to Higgs ``bag" formation, in as much as
the 1S and 2S states represent Higgs sources with a different degree of
localization, so they should be deformed differently by the formation of a
Higgs ``bag".  Our  strategy is to determine the ratio of the leading
perturbative correction to the 2S-1S splitting, to corrections that appear at
second order, after the feedback of the Higgs field on the toponium source wave
function has recorrected the energies of the toponium states. We would
interpret second order corrections to the splitting that were a significant
fraction of the first order correction, as evidence of a nonlinear feedback in
the Higgs-toponium system, representing the onset of ``bag" formation.

To examine the effect of  the interaction term, first consider as the zeroth
order approximation, a QCD toponium bound state. The energy level for the $n$S
state of such a system is given approximately by the Coulombic QCD binding
potential for heavy quarkonium
\begin{equation}
E_{nS}^{0}\simeq - \frac{4}{3} \frac{ (4 \pi \alpha_{s})^{2} m_{t}}{4 n^{2}}
\label{8.5}
\end{equation}
Here $\frac{4}{3}$ is the colour factor. For the 2S-1S splitting, $\Delta
E^{0}$ this gives $\Delta E^{0}\simeq -1.7$ GeV with our assumed value of the
effective QCD coupling. The modification of the splitting due to the presence
of the Higgs-top interaction is given by the change in the energy level
splitting, $\Delta E$, for which time independent non-degenerate perturbation
theory is used. The perturbing Hamiltonian is given by
\begin{equation}
{\cal H}_{1} = -  g \phi
\label{9}
\end{equation}
The first order correction to the  energy levels due to the presence of the
condensate $\phi$ is then:
\begin{equation}
E_{nS}^{1st} = < \psi_{nS}^{0} |- g \phi |\psi_{nS}^{0} >
\label{10}
\end{equation}
Applying  equation \ref{10} one can obtain numerical values for the first order
correction to the 2S-1S splitting for various values of $m_{H}$ and $m_{t}$.
Figure 1(a) shows the ratio of the first order correction to the 2S-1S
splitting to the zeroth order splitting, as a function of the top quark mass
and the mass of the Higgs. In Figure 1(b) contour lines are shown corresponding
to first order fractional shifts in the splitting of $5{\%}$ and of $1{\%}$;
 also shown on the figure is the top and Higgs mass parameter range allowed in
 the standard model by the constraints of vacuum stability, and perturbative
 non-triviality up to the Planck scale \cite{sher}. Clearly, a  measurable
 shift in the splitting from first order corrections is restricted to a small
 region of the allowed $m_{H}$ and $m_{t}$ parameter space .To test for
 evidence of ``bag" formation, one has to consider the higher order
 corrections  to the energy perturbation. In particular, Higgs ``bag" effects
 would be observable if the non-linear feedback in the Higgs-toponium system,
 represented by the second order correction, was large in comparison with the
 first order correction (say of the same order or more). A large second order
 correction implies that the fermion wave function is pulled in tighter,
 giving stronger binding to the toponium, and thereby indicating strong
 binding in a Higgs ``bag" potential well. This in turn would increase the
 influence of the source term in equation \ref{2}, and so result in a
 significant increase in the deviation of the Higgs field around the toponium
 which would then cause a further correction to the splitting. This nonlinear
 feedback would proceed to dig a hole in the Higgs field, and produce
 observable Higgs ``bag" effects.

Using the first order perturbations, and maintaining the top normalisation, one
has
\begin{equation}
\begin{array}{rcl}
\psi_{1S} &=& \psi^{0}_{1S} +  \psi^{1}_{1S} \\
\psi_{2S} &=& \psi^{0}_{2S} +  \psi^{1}_{2S} \nonumber
\end{array}
\label{11}
\end{equation}
for the first order corrected top wave functions. It should be noted that for
$m_{t}$ in the range 0 to 250 GeV the adjustment is slight. Given these
corrected wave functions, the correction to the Higgs field can be computed.
The correction to $\phi_{0}$
is given by
\begin{equation}
(\nabla^{2} - m_{H}^{2})\phi_{1} = g \psi^{0 \dagger}_{nS} \psi^{1}_{nS}
\label{13}
\end{equation}
As this equation only differs from equation \ref{2} in the inhomogenous term,
the Green's function is unaltered, and the $\phi_{1}$ can be found. This then
allows one to evaluate the second order correction to the toponium energy
levels. For the nS top wave functions, the second order energy correction is
\begin{equation}
E^{2nd}_{nS} = < \psi_{nS}^{0} |-g \phi_{0} | \sum_{m} b_{m} \psi_{mS}^{0} >
+ < \psi_{nS}^{0} |-g \phi_{1} | \psi_{nS}^{0} >
\label{14}
\end{equation}
where the $b_{m}$ are the  coefficients of the first order correction to the
wave function. The ratio of concern is
\begin{equation}
R =\frac{\Delta E^{2nd}}{\Delta E^{1st}} = \frac{ E^{2nd}_{2S}
-E^{2nd}_{1S}}{E^{1st}_{2S} -E^{1st}_{1S}}
\label{15}
\end{equation}
If R (plotted in Figure 2(a))is large then the feedback will have a significant
effect on the toponium bound state. On the other hand if R is negligible, then
then feedback is insignificant. Figure 2(b) displays the values of the mass
parameters required to give a $0.1{\%}$ and $1{\%}$ value for R. Clearly, the
 $m_{H}$ and $m_{t}$ for even such slight feedback are not physically
 acceptable as they lie outside the range of the  allowed mass parameters.
 Also, for any larger value of R the predicted values of $m_{H}$ and $m_{t}$
 fall further away from the acceptable region. The smallness of R in the mass
 parameter range allowed by the standard model tells one that the feedback
 corrections to the energy splitting are negligible, and thus Higgs ``bag" are
 experimentally unobservable. The only circumstance with marginally
 significant feedback is the case where the top mass is large ($m_{t} \approx
 150$ GeV) and the Higgs mass is of the order of a few Gev: this situation is
 already ruled out by LEP limits on the Higgs mass \cite{kar}.

If the Higgs sector is extended to a non-minimal content consisting  of two
Higgs doublets, then there will be extra physical scalar Higgs fields, each of
which must be considered. Consider the possibility that one or more of the
physical scalars in this non-minimal scenario has enhanced coupling to the top
quark. Such an enhancement will in general result in an increase in the
corrections to the energy level splittings, thus reopening the possibility for
detectable Higgs ``bag" effects. In Figure 3(a) the feedback ratio has been
plotted for a coupling that has been enhanced by a factor of 5 over the
standard model Higgs coupling, while Figure 3(b) indicates the mass parameters
required for R to reach the $1{\%}$ level. Clearly, while enhancement of the
 coupling increases the second order correction, even a factor of five
 increase in the coupling has not resulted in significant nonlinear feedback.
As such, we do not find evidence for Higgs ``bag" formation around toponium,
even with substantially enhanced couplings that could appear in models with
non-minimal Higgs content.

In conclusion, we have considered the possible observable effects of formation
of a Higgs ``bag" around toponium, as has been recently suggested. For values
of the Higgs and top mass expected in the standard model, the potentially
observable effects that could occur in toponium bound states are sufficiently
small, that no indication of non-linear feedback characteristic of ``bag"
formation has appeared. This conclusion remains essentially unaltered, even
with the ad hoc enhancement of the top-Higgs coupling by a factor of five, as
might occur in a model with a non-minimal Higgs sector.

\vskip .1in

\noindent{ {\bf Acknowledgements} } \\
\noindent  This work was supported in part by the
Natural Sciences and Engineering Research Council of Canada. One of us (ALM)
would like to thank C. Jaucent for encouragement and support during the course
of this research.

\newpage

\noindent{\bf{Figure Captions}}

\noindent Figure 1:(a) The ratio of the first order correction to the 2S-1S
splitting to the zeroth order splitting, as a function of the top quark mass
and the mass of the Higgs; (b) Contour lines corresponding to first order
fractional shifts in the splitting of $5{\%}$ (dashed line) and of $1{\%}$ (dot
 dashed line).

\noindent Figure 2: (a) The ratio R of the second order correction to the first
order correction for the 2S-1S energy splitting as a function of the top quark
mass and the mass of the Higgs;  (b) Contour lines  corresponding to $.1{\%}$
 (dot dashed line) and  $1{\%}$ (dashed line) in R.

\noindent Figure 3: (a) The R ratio for normal Higgs coupling (upper surface),
and for a coupling enhanced by a factor of 5 (lower surface) as a function of
the top quark mass and the mass of the Higgs;  (b) Contour line corresponding
to $1{\%}$ (dashed line) in R, with the enhanced coupling.

\end{document}